\documentstyle[12pt]{article}
\newcommand{\beq}{\begin{equation}}
\newcommand{\eeq}{\end{equation}}
\newcommand{\beqa}{\begin{eqnarray}}
\newcommand{\eeqa}{\end{eqnarray}}
\newcommand{\ba}{\begin{array}}
\newcommand{\ea}{\end{array}}

\begin{document}

\begin{flushright}
Preprint CAMTP/97-2\\
May 1997\\
\end{flushright}

\vskip 0.5 truecm
\begin{center}
\Large
{\bf  New universal aspects of diffusion in strongly chaotic systems}\\
\vspace{0.25in}
\normalsize
Marko Robnik, Jure Dobnikar, Andrea Rapisarda\dag,\\
Toma\v z Prosen\ddag \quad and Marko Petkov\v sek\S
\footnote{e--mails: robnik@uni-mb.si, jure.dobnikar@uni-mb.si,
andrea.rapisarda@ct.infn.it, prosen@fiz.uni-lj.si, 
marko.petkovsek@fmf.uni-lj.si}\\
\vspace{0.3in}
Center for Applied Mathematics and Theoretical Physics,\\
University of Maribor, Krekova 2, SLO--2000 Maribor, Slovenia\\
\dag INFN, Corso Italia 57, I-95129 Catania, Italy\\
\ddag Department of Physics, Faculty of Mathematics and Physics,\\
University of Ljubljana, Jadranska 19, SLO-1111 Ljubljana, Slovenia\\
\S Department of Mathematics, Faculty of Mathematics and Physics,\\
University of Ljubljana, Jadranska 19, SLO-1111 Ljubljana, Slovenia\\

\end{center}

\vspace{0.3in}

\normalsize
\noindent
{\bf Abstract.} We study some new universal aspects of diffusion
in chaotic systems, especially such having very large Lyapunov
coefficients on the chaotic (indecomposable, topologically transitive)
component. We do this by discretizing the chaotic component on 
the Surface-of-Section in a (large) number $N$ of simplectically 
equally big cells (in the sense of equal relative invariant ergodic 
measure, normalized so that the total measure of the chaotic 
component is unity). By iterating the transition of the chaotic 
orbit through SOS, where $j$ counts the number of iteration 
(discrete time), and assuming complete lack of correlations even
between consecutive crossings (which can be justified due to the very 
large Lyapunov exponents), we show the universal approach
of the relative measure of the occupied cells, denoted by $\rho(j)$,
to the asymptotic value of unity, in the following way:
$\rho(j) =  1 - (1-\frac{1}{N})^j$, so that in the limit of 
big $N$, $N\rightarrow \infty$, we have, for $j/N$ fixed, the exponential law
$\rho(j) \approx 1 - \exp (-j/N)$. This analytic result is verified
numerically in a variety of specific systems: For a plane
billiard (Robnik 1983, $\lambda=0.375$), for a 3-D billiard (Prosen 1997,
$a=-1/5, b=-12/5$), for ergodic logistic map (tent map), for
standard map ($k=400$) and 
for hydrogen atom in strong magnetic field ($\epsilon=-0.05$)
the agreement is almost perfect (except, in the latter two systems, 
for some long-time deviations on
very small scale), but for H\'enon-Heiles system ($E=1/6$) and
for the standard map  ($k=3$) the deviations are noticed although
they are not very big (only about 1\%).
We have tested the random number generators (Press et al 1986), 
and confirmed that some are almost perfect (ran0 and ran3),
whilst two of them (ran1 and ran2) exhibit big deviations (in units
of the theoretically expected standard deviation, as much as 20). 
Thus physical deterministic chaotic systems can be better simulators
of random numbers than some well known mathematical algorithms.
We give an outline
of an improved analytical theoretical model (the so-called two and many
component model), where deviations from the exponential law can
be captured in a statistical way.

\vspace{0.6in}

Submitted to {\bf Journal of Physics A: Mathematical and General}
\normalsize
\vspace{0.1in}
  
\newpage

%\section{Introduction}

\noindent
One of the major open problems in the mathematics of nonlinear
Hamiltonian (conservative) chaotic systems of the KAM type is the proof of the
so-called coexistence problem (Strelcyn 1991), i.e. the proof that the chaotic
components have positive measure. (The KAM Theorem guarantees
that the set of invariant tori has positive measure, whose complement
is small with the perturbation parameter (Kolmogoroff 1954, 
Arnold 1963, Moser 1962, Benettin et al 1984, Gutzwiller 1990).) The chaotic
component could be defined e.g. by the positivity of the (largest)
Lyapunov exponent, which is a sufficient but not a necessary 
criterion\footnote{For example, in nonrational plane polygonal billiards all
Lyapunov exponents are strictly zero (Sinai 1976), and yet they
can be ergodic.}.
We shall define a chaotic component as the closure of a dense
chaotic orbit, which is thus assumed to be an
indecomposable invariant component (topologically transitive, i.e.
containing a chaotic dense orbit).
\\\\
There are really no serious doubts about the positivity of the
measure of the chaotic components, and so in physics we rely on
heuristic arguments to actually assume positivity. Then the 
question is how to calculate the symplectic (invariant and ergodic)
measure of the chaotic component.
\\\\
We have approached this problem in a recent extensive work (Dobnikar 1996,
Robnik and Dobnikar 1997) on the dynamics in a plane billiard system,
defined as the quadratic conformal map of the unit disc (in the complex
$z$-plane) onto the physical (complex) $w$-plane, $w(z) = z +\lambda z^2$.
This system has been introduced by Robnik (1983), and recently 
extensively studied for many different values of $\lambda$ by many authors, 
in a variety of contexts and even in experimental setups like quantum dots
(Bruus and Stone 1994, Stone and Bruus 1993ab), optical model 
(N\"ockel and Stone 1997, N\"ockel et al 1996), 
microwave cavities (Rehfeld et al 1996, Richter 1996,
St\"ockmann et al 1997). 
Further dynamical details were corrected in (Hayli et al 1987).
\\\\
At $\lambda=0$ we have the integrable 
case of the circle billiard, for $0 < \lambda <1/4$ the billiard
is convex, and since the boundary is analytic, the KAM theory applies
(Lazutkin 1981, 1991), at $\lambda = 1/4$ we get the first point 
of zero curvature at the boundary point $w=w(z=-1)=-3/4$, allowing for
the breaking of Lazutkin caustics (invariant tori associated with
the boundary glancing orbits), and for $1/4 < \lambda <1/2$ the billiard
is non-convex, largely and strongly chaotic (very tiny islands
of stability), probably becomes rigorously ergodic at some 
$\lambda \ge 0.2775$ (Li and Robnik 1994), and is definitely
proven to be rigorously ergodic for $\lambda = 1/2$ (the so-called
cardioid billiard, having the cusp singularity at $w=w(z=-1)=-1/2$) 
(Markarian 1993). The cardioid billiard has been studied also by
B\"acker, Steiner and Stifter (1995) and  by B\"acker and Dullin (1997).
\\\\
Our main problem was to calculate numerically, accurately and reliably,
the measure of chaotic components. Working in the KAM regime
($0 < \lambda < 1/4$) we were observing the typical KAM hierarchy of
smaller and smaller islands of stability surrounded by chaotic
components, the details of which will be reported in a separate
paper (Robnik and Dobnikar 1997), but are reported already in 
(Robnik 1983).

The main objective was to calculate the fractional measure 
(the relative area on the SOS, the latter being defined by the 
Poincar\'e-Birkhoff coordinates) of the largest chaotic component at 
given $\lambda$, which we traditionally denote by $\rho_2$.

This parameter is important also in treating the related quantum
mechanical problem (solutions of the Helmholtz equation with the Dirichlet
boundary conditions on the billiard boundary) (Berry and Robnik 1984,
Prosen and Robnik 1993,1994a-b, Li and Robnik 1995).

We have discovered, to our big surprise, that this numerical calculation
is extremely difficult, and as one consequence, some of the previous
results had to be revised. By dividing the SOS into a large number
of rectangular grid cells of equal relative (normalized) measure we 
have calculated the relative  measure of the
chaotic component $\rho_2$ by three different methods: 
{\bf (M1)} calculating the
Lyapunov exponent for a trajectory starting in a given cell and summing up
the area of cells having the positive Lyapunov exponent; 
{\bf (M2)} calculating
two nearby trajectories, separated by an infinitesimal distance
(e.g. single precision e-8 while all calculations were in double precision
e-16) and summing the cells exhibiting macroscopic divergence in a reasonable
time; 
and {\bf (M3)} starting the chaotic trajectory and counting
(summing up) the area of the cells visited ("black cells"). 

It turned
out that the third method {\bf (M3)} is the best, fastest, and most reliable.
To improve the result one has to enlarge the number of cells $N$ 
on the chaotic component. But then, the time $j$ (of iterating the
map on the SOS, the discrete time) has to be taken much larger
(by at least several orders of magnitude) than $N$, otherwise
the statistics of visiting cells (black cells) would be insignificant.
Even after calculating as many as $10^9$ iterations, with $N=1000-2000$,
 the result was
no better than within 1\%. One reason is that the boundary of
the chaotic component turned out to have relatively large fractal dimension
around 1.56.  The difficulties of estimating the asymptotic
value of the relative area $\rho_2$ of the chaotic component led us to the
careful investigation of the evolution of relative area of occupied 
(black) cells $\rho_2(j)$ as time $j$ proceeds, and we wanted 
to understand that theoretically, in order to be able to make better estimates 
of the limiting asymptotic value  $\rho_2 \equiv \rho_2(j=\infty)$.
From here onwards we shall use the notation 
$\rho(j) \equiv \rho_2(j)/\rho_2(j=\infty)$.
\\\\
In general this time evolution $\rho(j)$ with $j$ is quite
complex and specific, nonuniversal, depending on many features
appearing in the phase space (SOS), e.g. existence of sticky
objects like cantori can affect a temporary but quite persistent
trapping of the orbit near such an object, which is then
manifested in a transient plateau of the curve $\rho(j)$,
which sometimes might be mistakenly interpreted as final and definite
convergence of the cumulative area/volume $\rho_2(j)$. However,
if the system is really strongly chaotic, having large maximal
positive Lyapunov exponent, then due to the bound motion and
conservation of the phase space volume, we find a very strong
stretching and folding in the phase space. In such a limiting
case therefore one small phase space cell (SOS cell) becomes uniformly
 distributed, in the coarse grained sense, all over the allowed 
chaotic component.
Thus, in such extreme case, the probability of entering 
a given cell belonging to the same chaotic component in SOS is just 
equal to the relative measure
of the cell, i.e. there are no correlations, not even between two
consecutive SOS iterations: complete randomness of deterministic
motion.
\\\\
While the behaviour in such ideal extreme case is quite obvious,
it is far from obvious that the conditions of the complete
randomness are actually satisfied in specific chaotic deterministic 
dynamical systems. Therefore, to make things precise we have developed
the following theoretical model. 
\\\\
Suppose we have $N$ cells, where their order and geometry of
arrangement is completely irrelevant and perhaps
not even known. We are filling the cells with balls, one at the time.
At each step $j$ (discrete time) we have equal probability
to choose any cell, equal to $a\equiv 1/N$. (Thus there
are absolutely no correlations between any moves, 
including the consecutive ones, and therefore e.g. the repetition of
falling into  a given already filled cell is allowed.) We define by
$P_j(k)$ the probability that at $j$-th step $k$ cells are occupied,
keeping in mind that $a=1/N$ is the model parameter implicit in
the mathematical formulae. We shall refer to this model as 
{\em the random model} (of strongly chaotic deterministic diffusion).
The probabilities must be normalized, therefore

\beq
\sum_{k=1}^{k=N} P_j(k)=1, \qquad \forall j=1,2,\dots.
\label{eq:normalize}
\eeq
We shall calculate  $P_j(k)$, and their moments, in particular
the first moment, namely the average (normalized) measure of the
occupied cells, 

\beq
\rho(j) \equiv \sum_{k=1}^{N} ka P_j(k) = \langle ka \rangle,
\label{eq:defrho}
\eeq
where  by $\langle...\rangle$ we denote the averaging operation.
\\\\
Before explicitly calculating $P_j(k)$ we observe the physically
(probabilistically) quite obvious recursion relation, namely

\beq
P_{j+1}(k+1) = P_j(k+1) \frac{k+1}{N} + P_j(k) (1- \frac{k}{N}),
\qquad \forall 0\leq k\leq j,
\label{eq:recursion}
\eeq
where we also define the boundary conditions

\beq
P_j(0)\equiv 0, \qquad P_1(1)=1,  \qquad P_j(k>j)=0, \qquad P_j(k>N)=0.
\label{eq:boundarycond}
\eeq
The interpretation of equation (\ref{eq:recursion}) is: the 
probability to have $(k+1)$ cells occupied at time $(j+1)$ is
equal to the sum of the following probabilities: either
at time $j$ the $(k+1)$ cells were already occupied, and we
add the next ball into the black cells with probability $(k+1)/N$,
or at time $j$ only $k$ cells are occupied (black), and we add the next 
ball (the $(j+1)$-st one) into the empty (not-yet-occupied) cells
with probability  $(1-k/N)$. With the boundary conditions
(\ref{eq:boundarycond}) the recursion relation  (\ref{eq:recursion})
solves the problem, in principle. We show the explicit solution
below, using different approach. However, for a practical
(numerical) evaluation of $P_j(k)$ (on the computer) it is much 
better to use the recursion formula (\ref{eq:recursion}) than
the explicit result which we shall derive below.
\\\\
For the beginning please observe that the summation of the
recursion equation (\ref{eq:recursion}) on each side from
$k=0$ to $k=N-1$ confirms the preservation of normalization
(\ref{eq:normalize}), for all $j$.
\\\\
Next, we can find the solution for $\rho(j)$ at once by the
following trick: multiply the recursion relation (\ref{eq:recursion})
on the left and on the right by $(k+1)/N$ and sum it from $k=0$ to
$k=N-1$. By denoting $S(j)$ the second moment,

\beq
S(j) \equiv \langle (ak)^2 \rangle = \sum_{k=1}^{k=N} P_j(k) (ak)^2,
\label{eq:defS}
\eeq
we obtain in a straightforward manner,

\beq
\rho(j+1) = S(j) - (S(j)-P_j(N)) +(1-a)(\rho(j) -P_j(N))
+a(1-P_j(N)),
\label{eq:mom1}
\eeq
and therefore after cancellation of $S(j)$'s we get the simple
recursion equation for $\rho(j)$, namely

\beq
\rho(j+1)=a + (1-a)\rho(j),
\label{eq:rhorec}
\eeq
with the explicit solution, quite easy to find,

\beq
\rho(j) = 1 - (1-a)^j = 1 - (1- \frac{1}{N})^j,
\label{eq:solrho}
\eeq
which in the limit of large $N$, for $j/N$ fixed,
becomes simple exponential law

\beq
\rho(j) \approx 1 - \exp(-\frac{j}{N}).
\label{eq:exponlaw}
\eeq
We see that in the limit of sufficiently large $N$, for $j/N$ fixed, 
we have the universal scaling of the relative measure of chaotic region
$\rho$, normalized to unity,
such that $\rho(j) \rightarrow 1$, when $j\rightarrow \infty$,
 as a function of the scaled discrete time, namely $j/N$.
We shall show and see below, that this law is obeyed by a surprising
variety of deterministic dynamical systems.
\\\\
Our {\em random model} is a probabilistic (statistical) model and
therefore we can calculate all moments of $P_j(k)$, systematically,
using the same trick as above: multiply the recursion relation
(\ref{eq:recursion}) by $(k+1)^2/N^2$ on both sides, sum it up
from $k=0$ to $k=N-1$ on both sides, and uncover the recursion
relation for the second moment $S(j)$, namely

\beq
S(j+1) = \frac{1}{N^2} + (\frac{2}{N} - \frac{1}{N^2})\rho(j) 
+ (1-\frac{2}{N}) S(j),
\label{eq:Srec}
\eeq
and by using the exact result for $\rho(j)$ from equation 
(\ref{eq:solrho}) we have

\beq
S(j+1) = 2a -a(2-a)(1-a)^j + (1-2a)S(j), \qquad a\equiv 1/N.
\label{eq:Srecf}
\eeq
This equation can be solved either by standard technique or
by using the definition of $S(j)$, equation (\ref{eq:defS}),
to yield the explicit result

\beq
S(j) = 1 - (2-a)(1-a)^j + (1-a)(1-2a)^j, \qquad a\equiv 1/N,
\label{eq:solS}
\eeq
so that the predicted dispersion $\sigma^2(j)$ according 
to our model is exactly

\beq
\sigma^2(j) = S(j) - \rho^2(j) = a(1-a)^j+(1-a)(1-2a)^j-(1-a)^{2j},
\quad a\equiv 1/N.
\label{eq:solsigma}
\eeq
In the asymptotic limit of sufficiently large number of cells
$N=1/a \rightarrow \infty$, but keeping $j/N$ fixed, 
we find the simple exponential laws:

\beq
\rho(j)\approx 1-\exp(-j/N), \qquad S(j)\approx [1-\exp(-j/N)]^2 \approx
\rho^2(j),
\label{eq:asymptrhoS}
\eeq
and therefore
\beq 
\sigma^2(j) \approx N^{-1} \lbrack \exp(-j/N)- \exp(-2j/N)\rbrack \rightarrow 0.
\label{eq:asymplaws}
\eeq

\vspace{0.3in}
\noindent
Now we show the explicit and exact result for $P_j(k)$, just for
the sake of completeness. In fact the quantity we seek is the subject of 
the classical problem from combinatorial analysis, treated 
e.g. in (Vinogradov 1979, Vol.2, p.973, Riordan 1978, p.48, Table 2,
Graham, Knuth and Patashnik 1994): 
The question is how many possibilities are there to 
distribute $j$ different things into $N$ different cells
under the condition that $N-k$ cells are empty (i.e. precisely  
$k$ cells are occupied): The answer is well known,
namely in the literature denoted by $C_{Nj}(N-k)$,

\beq
C_{Nj}(N-k) =  {N \choose N-k} k! {\cal S}(j,k) = \frac{N!}{(N-k)!}
{\cal S}(j,k),
\label{eq:comb}
\eeq
where ${\cal S}(j,k)$ are the so-called Stirling numbers of the
second kind (Vinogradov 1979, Riordan 1978, Graham, Knuth and Patashnik 1994)

\beq
{\cal S}(j,k) \equiv \frac{1}{k!} \sum_{i=0}^{k} 
{k \choose i}i^j(-1)^{k-i},
\label{eq:stirling2}
\eeq
which are known to satisfy the triangular recursion relation
${\cal S}(j,k) = k {\cal S}(j-1,k) + {\cal S}(j-1,k-1)$, where
${\cal S}(0,0) =1, \quad {\cal S}(j,0)=0, \quad {\cal S}(0,k)=0 
\quad (j,k)>0$.
Of course, having the $N$ cells and $j$ things, like in our
{\em random model},  the total
number of possibilities to distribute $j$ things (balls) into $N$
cells is just $N^j$, and therefore we have the final and complete explicit
solution to the {\em random model}, namely

\beq
P_j(k) = \frac{C_{Nj}(N-k)}{N^j} = \frac{N! {\cal S}(j,k)}{(N-k)! N^j}.
\label{finsolP}
\eeq

\vspace{0.3in}
\noindent
Now we proceed by analyzing specific dynamical systems from the point
of view of the statistical theory  presented in our {\em random model},
to see to what extent we find agreement in real systems. The really
big surprise is that the behaviour was found in excellent
or even perfect agreement with theory 
in a large variety of deterministic dynamical
systems, sufficiently far from a pronounced KAM-regime, by which we mean
either close to ergodic (big $\rho_2(j=\infty) \approx 1$), or strongly
chaotic (big Lyapunov exponent but not necessarily very 
large $\rho_2(j=\infty)$).
\\\\
We have investigated the plots $\rho(j)$ versus $j/N$ for the following
systems: 
{\bf (a)} the 2-D billiard (Robnik 1983) at $\lambda=0.375$,
{\bf (b)} the 3-D billiard (Prosen 1997, def. geom.: $a=-1/5, b=-12/5$), 
{\bf (c)} the hydrogen atom in strong magnetic field (DKP = diamagnetic Kepler
problem) with the {\em scaled energy} $\epsilon=-0.05$ (see
e.g. Hasegawa, Robnik and Wunner 1989), 
{\bf (d)} the H\'enon-Heiles system
at the escape (dissociation) energy $E=1/6$ (see H\'enon and Heiles
1964), 
{\bf (e)} the standard map (Chirikov 1979) with $k=3$,
{\bf (f)} the standard map (Chirikov 1979) with $k=400$, and
{\bf (g)} for the logistic map at $\lambda=4$ (the ergodic tent map).

In case of smooth systems {\bf (c)} and {\bf (d)} we have used
special symplectic integration routines, devised by Yoshida (1990).
This enabled a fast and extremely accurate calculations, allowing
us to compute about the order of magnitude $10^5$ iterations on
the SOS.
\\\\
The agreement for all systems was perfect (deviations much smaller
than 1\%), except in {\bf (d)} and {\bf (e)}, where the deviations
fluctuated around up to 1\%. Therefore we do not show these plots,
since all curves practically overlap with the theoretical
curve (\ref{eq:solrho}) and ({\ref{eq:exponlaw}) within the graphical
resolution. The small deviations
stem from the fact that the Lyapunov coefficients are still not big
enough, and also that there might be significant episodes of transient
behaviour in the relationship of $\rho(j)$ with respect to the
discrete time $j/N$. Such transient episodes are typically caused
by the sticky objects in the phase space, e.g. by cantori, where
the classical orbit spends long time before resuming the chaotic
random filling of the remaining empty cells. They are very well
manifested in systems with more pronounced KAM structure, like
e.g. the 2-D billiard ($\lambda=0.15$) or 3-D billiard (for sufficiently
small $a$ and $b$, such as $a=-0.1$, $b=0$).
In order to describe
such systematic effects in a statistical way we have developed a
{\em multicomponent random model}, where orbital transitions inside each
component are random as in our {\em random model}, however, they
might jump (rarely) from one into another component. Some of
the dynamical features are well described by such a model, 
whose detailed description will be published in a separate paper
(Robnik, Prosen and Dobnikar 1997).
\\\\
In cases {\bf (a)} - {\bf (c)} the agreement is so good, quite
surprisingly, that it is necessary to magnify the scale so
that the details of deviations become clearly visible. This is
done in figures 1(a-c) for the systems, respectively.
We plot the fluctuation (difference) $\rho_{numerical}(j) -
\rho_{theory}(j)$ (the noisy curves) to be compared with  the
theoretically expected standard deviation 
$\sigma(j) = \pm\sqrt{S(j) - \rho^2(j)}$ (smooth
curves),  again as a function of the scaled discrete
time $j/N$.  We do this in each of the plots for
one initial condition and for the average over 50 randomly
(uniformly over the chaotic component) chosen initial conditions
which suppress the dispersion by a factor of 50, of course,
and the standard deviation by $\sqrt{50}$.
\\\\
The conclusion in inspecting  these plots is that the 2-D billiard
 perfectly obeys the law of our {\em random
model}, whilst for the hydrogen atom in a strong magnetic 
field and the 3-D billiard
we uncover systematic deviations from the theory for
sufficiently large scaled discrete times $j/N$: the deviations are
orders of magnitude bigger than the prediction of our {\em random model}
for the DKP, but somewhat smaller in the 3-D billiard.
It should be noted that the deviations are almost strictly {\em negative},
reflecting the fact that the physical orbits like to stick to
already occupied cells (existence of sticky objects in the phase space).
\\\\
It is interesting to look at the results for random number
generators. Cells are "randomly" chosen, using different 
random number generator algorithms. One such algorithm was 
devised by Finocchiaro et al
(1993) in the context of nuclear physics. The agreement was
perfect and so we do not show the fluctuations plots.

Also, we have checked and tested some well known algorithms
for random number generators, such as ran0, ran1, ran2 and ran3
devised and described in (Press et al 1986, ch. 7, and the references
therein). The results are shown in figure 2. As we see ran0 and
ran3 are in excellent agreement  
with our {\em random model}, whilst
for ran1 and ran2 random generators the deviations become clearly
very large. Thus,  for example, our dynamical deterministic 2-D 
billiard systems, or the logistic map, or the standard map at $k=400$, 
are  in fact better random number generators than some of the
most familiar random number generators used in the computers.
\\\\
Finally, we have looked at the irrational triangle (the
angles are $\alpha=(\sqrt{5}-1)\pi/4$, $\beta=(\sqrt{2}-1)\pi/2$) as one
ergodic system with strictly zero Lyapunov exponents (Sinai 1976).
Even there the agreement with our {\em random model} is surpringly
good on the largest scale (figure 3), while the fluctuation
diagramme exhibits large deviations, again negative, showing that
the real billiard orbits like to stay in the already occupied 
regions (sticky objects in phase space).
\\\\
In conclusion, we have developed {\em the random model} of
stochastic diffusion of dynamical systems with invariant measure
on their Surfaces of Section, which is supposed and confirmed
to apply very well in strongly chaotic systems (for which
the Lyapunov coefficients on the chaotic component are sufficiently
large). We have discovered and explained the universal scaling
behaviour of the normalized chaotic measure $\rho$  as 
a function of the scaled discrete time $j/N$, where $N$ is the
number of cells: namely, in the limit of sufficiently large $N$, 
for $j/N$ fixed, we have 
a simple exponential law $\rho(j) = 1 - \exp(-j/N)$. We also
predict the higher moments, especially the dispersion given
in equation (\ref{eq:solsigma}). The model is solved
completely in the sense that we have calculated the probabilities
$P_j(k)$ of having exactly $k$ non-empty cells at time $j$.
Therefore, one can calculate all the moments.
The deviations from the predictions of the random model are
qualitatively understood, but will be treated in detail
together with a new, more general theory (the multicomponent model) in
a separate work (Robnik, Prosen and Dobnikar 1997).
The work is in a sense extension of the theory
of transport in Hamiltonian systems by MacKay, Meiss and Percival
(1984).

\section*{Acknowledgements}
\par
The financial support by the Ministry of Science
and Technology of the Republic of Slovenia is acknowledged with
thanks. A.R. thanks INFN for financial support.

\newpage

\section*{References} 
\parindent=0. pt

Arnold V I 1963 {\em Usp. Mat. Nauk SSSR} {\bf 18} 13
\\\\
B\"acker A, Steiner F and Stifter P 1995 {\em Phys.  Rev. E} {\bf 52} 2463
\\\\
B\"acker A and Dullin A 1997 {\em J.Phys.A: Math.Gen.} {\bf 30} 1991
\\\\
Benettin G C, Galgani L, Giorgilli A and Strelcyn J.-M. 1984
{\em Nuovo Cimento B} {\bf 79} 201
\\\\
Berry M V and Robnik M 1984 {\em J.Phys.A: Math.Gen.} {\bf 17} 2413
\\\\
Bruus H and Stone A D 1994 {\em Phys.Rev. B} {\bf 50} 18 275
\\\\
Chirikov B V 1979 {\em Phys.Rep.} {\bf 52} 263
\\\\
Dobnikar J 1996 {\em Diploma Thesis}, Department of Physics,
University of Ljubljana and CAMTP University of Maribor, October 
1996, in Slovenian, unpublished
\\\\
Finocchiaro P, Agodi C, Alba R, Bellia G, Coniglione R, Del Zoppo A,
Maiolino C, Migneco E, Piattelli P and Sapienza P 
1993 {\em Nucl. Instrum. Methods.} {\bf 334} 504 
\\\\
Graham R L, Knuth D E and Patashnik O 1994 {\em Concrete Mathematics} 
(Reading, Mass.: Addison-Wesley) p.257
\\\\
Gutzwiller M C 1990 {\em Chaos in Classical and Quantum Mechanics}
(New York: Springer) p.132
\\\\
Hasegawa H, Robnik M and Wunner G 1989 {\em Prog. Theor. Phys. Suppl. 
(Kyoto)} {\bf 98} 198
\\\\
Hayli A, Dumont T, Moulin-Ollagier J and Strelcyn J.-M. 1987 
{\em J.Phys.A: Math.Gen.} {\bf 20} 3237
\\\\
H\'enon M and Heiles C 1964 {\em Astron.J} {\bf 69} 73
\\\\
Kolmogoroff A N 1954 {\em Dokl. Akad. Nauk SSSR} {\bf 98} 527
\\\\
Lazutkin V F 1981 {\em The Convex Billiard and the Eigenfunctions of the 
Laplace Operator} (Leningrad: University Press) in Russian
\\\\
Lazutkin V F 1991 {\em KAM Theory and Semiclassical Approximations
to Eigenfunctions} (Heidelberg: Springer)
\\\\
Li Baowen and Robnik M 1994 to be published
\\\\
Li Baowen and Robnik M 1995 {\em J.Phys.A: Math.Gen.} {\bf 28} 4483 
\\\\
MacKay R S, Meiss J D and Percival I C 1984 {\em Physica } {\bf 13D} 55
\\\\
Markarian R 1993 {\em Nonlinearity} {\bf 6} 819
\\\\
N\"ockel J U and Stone A D 1997 {\em Nature} {\bf 385} 45
\\\\
N\"ockel J U, Stone A D, Chen G, Grossman H and Chang R K 1996
 {\em Opt. Lett.} {\bf 21} 451609
\\\\
Moser J 1962 {\em Nachr. Akad. Wiss. G\"ottingen} 1
\\\\
Press W H, Flannery B P, Teukolsky S A and Vetterling W T 1986
{\em Numerical Recipes} (Cambridge: cambridge University Press) Ch.7
\\\\
Prosen T 1997a,b {\em Phys.Lett.A} in press
\\\\
Prosen T and Robnik M 1993 {\em J.Phys.A: Math.Gen.} {\bf 26} 2371
\\\\
Prosen T and Robnik M 1994a {\em J.Phys.A: Math.Gen.} {\bf 27} L459
\\\\
Prosen T and Robnik M 1994b {\em J.Phys.A: Math.Gen.} {\bf 27} 8059
\\\\
Rehfeld H, Alt H, Dembowski C, Gr\"af H.-D., Hofferbert R, Lengeler H and
Richter A 1996 {\em Wave Dynamical Chaos in Superconducting
Microwave Billiards}, Preprint to appear in the Proc. of the
3rd Int'l Summer School/Conference {\em Let's Face Chaos through
Nonlinear Dynamics}, held at the University of Maribor on 24 June- 5 July 
1996, to be published in {"Open Systems and Information Dynamics"} 1997
\\\\
Richter A 1996 {\em "Playing Billiards with Microwaves - Quantum
Manifestations of Classical Chaos"}, to appear in Proc. of the
Workshop {\em "Emerging Applications of Number Theory"}, University
of Minnesota, Mineapolis.
\\\\
Riordan J 1978 {\em An Introduction to Combinatorial Analysis}
(Princeton: Princeton University Press) p.32
\\\\
Robnik M 1983 {\em J.Phys.A:Math.Gen.} {\bf 16} 3971
\\\\
Robnik M and Dobnikar J 1997 to be published
\\\\
Robnik M, Prosen T and Dobnikar J 1997 to be published
\\\\
Sinai Ya G 1976 {\em Introduction to Ergodic Theory} (Princeton: Princeton
University Press) p.140
\\\\
Stone A D and Bruus H 1993a {\em Physica} {\bf 189B} 43
\\\\
Stone A D and Bruus H 1993b {\em Surface Science} {\bf 305} 490
\\\\
St\"ockmann H.-J., Kuhl U, Robnik M, Dobnikar J and Veble G 1997,
work in progress
\\\\
Strelcyn J.-M. 1991 {\em Colloquium Mathematicum} {\bf LXII} Fasc.2 
331-345
\\\\
Vinogradov I M (Editor) 1979 {\em Mathematical Encyclopaedia} (Moscow: Soviet. 
Encycl.) Vol. 2, p.971, in Russian
\\\\
Yoshida H 1990 {\em Phys.Lett.A} {\bf 150} 262
\newpage

\section*{Figure captions}

\vspace{0.3in}
\noindent
{\bf Figure 1(a-c)}: We show the plots of $\rho_{numerical}(j) - 
\rho_{theory}(j)$ versus the scaled discrete time $j/N$. 
The outer noisy curve is the numerical result for a chaotic orbit
with a certain representative initial condition, whilst the inner one
is the average over fifty evenly distributed initial conditions.
We also show the $\pm \sigma(j)$ standard deviation, as predicted 
theoretically in equation (13). It is clearly seen that for the
2-D billiard in (a) the agreement is almost perfect, in the sense
that the fluctuations are within the predicted range, whilst
for the 3-D billiard in (b) we see systematic deviations, obviously
caused by some perhaps unexpected long time correlations. 
Such correlations are even stronger in case of the hydrogen atom in 
a strong magnetic field, shown in (c). In all cases the deviations
are predominantly negative: the orbit tries to stick to some of the already
occupied cells.

\vspace{0.3in}
\noindent
{\bf Figure 2}: We show the results for the random number generators,
compared by the theoretical $\pm\sigma(j)$ curves according to
equation (13) (Press et al 1986): for ran0 and ran3 the agreement
is excellent, while for ran1 and ran2 the deviations are so big,
that they are actually not random: they seem to repel from the
occupied cells.

\vspace{0.3in}
\noindent
{\bf Figure 3}: We show the global plot, $\rho_{numerical}(j)$ (full line)
and $\rho_{theory}(j)$ (dashed), for the
irrational triangle, where the agreement is surprinsingly good,
in spite of strictly vanishing Lyapunov exponents. In
the fluctuations diagramme we plot 
$\rho_{numerical}(j) - \rho_{theory}(j)$ versus $j/N$, together
with the $\pm\sigma$, where we see the same trend to negative
deviations due to the sticky objects in the phase space.

\end{document}